\documentclass[onecolumn]{aastex61}
\usepackage{amsmath,amstext}
\usepackage{hyperref}
\usepackage[T1]{fontenc}
\usepackage{apjfonts} 
\usepackage[figure,figure*]{hypcap}
\DeclareMathOperator{\arcsinh}{arcsinh}
\newcommand{\asyr}{$''\,\mathrm{yr}^{-1}$}
\newcommand{\masyr}{$\mathrm{mas}\,\mathrm{yr}^{-1}$}
\shorttitle{Backyard Worlds: Planet 9}
\shortauthors{Kuchner et al.}

\begin{document}

\title{The First Brown Dwarf Discovered by the Backyard Worlds: Planet 9 \\ Citizen Science Project}

\author[0000-0002-2387-5489]{Marc J. Kuchner}
%\email{Marc.Kuchner@nasa.gov}
\affil{NASA Goddard Space Flight Center, Exoplanets and Stellar Astrophysics Laboratory, Code 667, Greenbelt, MD 20771}
\author[0000-0001-6251-0573]{Jacqueline K. Faherty}
\affiliation{Department of Astrophysics, American Museum of Natural History, Central Park West at 79th St., New York, NY 10024, USA}
\author[0000-0001-5106-1207]{Adam C. Schneider}
\affil{School of Earth and Space Exploration, Arizona State University, Tempe, AZ, 85282, USA}
\author[0000-0002-1125-7384]{Aaron M. Meisner}
\affil{Berkeley Center for Cosmological Physics and Lawrence Berkeley National Laboratory, Berkeley, CA 94720, USA}
\author[0000-0002-0201-8306]{Joseph C. Filippazzo}
\affil{Space Telescope Science Institute, 3700 San Martin Dr., Baltimore, MD 21218, USA}
\author[0000-0002-2592-9612]{Jonathan Gagn\'e}
\affiliation{Carnegie Institution of Washington DTM, 5241 Broad Branch Road NW, Washington, DC~20015, USA}
\affiliation{NASA Sagan Fellow}
\author[0000-0002-1113-4122]{Laura Trouille}
\affil{Adler Planetarium, 1300 S. Lake Shore Drive, Chicago, IL 60605}
\author[0000-0002-3741-4181]{Steven M. Silverberg}
\affil{NASA Goddard Space Flight Center, Exoplanets and Stellar Astrophysics Laboratory, Code 667, Greenbelt, MD 20771}
\author{Rosa Castro}
\affil{Touro Worldwide University, Organizational Psychology Department}
\author{Bob Fletcher}
\affil{Lambert School, 38 Church Street, North Hobart, Tasmania, Australia 7000}
\author[0000-0002-7781-3339]{Khasan Mokaev}
\affil{Russian Federation, Kabardino-Balkar Republic,  Babugent}
\author[0000-0002-1799-5783]{Tamara Stajic}
\affil{School of Electrical Engineering, University of Belgrade, Bulevar kralja Aleksandra 73, 11120 Belgrade, Serbia}

\begin{abstract}
The {\it Wide-field Infrared Survey Explorer} ({\it WISE}) is a powerful tool for finding nearby brown dwarfs and searching for new planets in the outer solar system, especially with the incorporation of NEOWISE and NEOWISE-Reactivation data. So far, searches for brown dwarfs in {\it WISE} data have yet to take advantage of the full depth of the {\it WISE} images. To efficiently search this unexplored space via visual inspection, we have launched a new citizen science project, called ``Backyard Worlds: Planet 9,'' which asks volunteers to examine short animations composed of difference images constructed from time-resolved {\it WISE} coadds. We report the discovery of the first new substellar object found by this project, WISEA J110125.95$+$540052.8, a T5.5 brown dwarf located approximately 34\,pc from the Sun with a total proper motion of $\sim$0.7\,\asyr. WISEA J110125.95$+$540052.8 has a {\it WISE} W2 magnitude of $W2=15.37 \pm 0.09$; this discovery demonstrates the ability of citizen scientists to identify moving objects via visual inspection that are 0.9 magnitudes fainter than the W2 single-exposure sensitivity, a threshold that has limited prior motion-based brown dwarf searches with {\it WISE}.

\end{abstract}


%-------------------------------------------------%
\section{Introduction}
\label{sec:intro}
%-------------------------------------------------%

Citizen science uses the power of numerous volunteers to tackle research projects that would otherwise be impossible or impractical to accomplish. This mode of research has deep roots in the field of astronomy, epitomized by the American Association of Variable Star Observers (AAVSO), founded in 1911. Now, through the Zooniverse community, citizen scientists perform an array of astronomical tasks online, like classifying galaxy morphologies (e.g., \citealt{2008MNRAS.389.1179L}) analyzing and discovering transiting exoplanets (e.g., \citealt{2012ApJ...754..129S}, \citealt{2013ApJ...768..127S}), and sorting circumstellar disk candidates from false positives \citep{2016ApJ...830...84K}.  Some of these projects have led to truly remarkable discoveries, like ``Hanny's Voorwerp'', a region of bright emission near a spiral galaxy thought to be due to a quasar light echo \citep{2009MNRAS.399..129L}, KIC 8462852 (aka ``Tabby's Star''), a star showing large, irregularly shaped dips in its light curve that have not yet been fully explained \citep{2016MNRAS.457.3988B}, and AllWISE J080822.18--644357.3, a remarkably old (45\,Myr) accreting M dwarf hosting a pre-transitional disk \citep{2016ApJ...830L..28S,2017arXiv170304544M}.  In addition to astronomy projects, Zooniverse currently hosts over fifty projects across the disciplines from climate science to history.  

The {\it Wide-field Infrared Survey Explorer} \citep[{\it WISE};][]{2010AJ....140.1868W} surveyed the whole sky simultaneously in four bands with central wavelengths of 3.4, 4.6, 12, and 22\,$\mu$m (hereafter referred to as {\it W1}, {\it W2}, {\it W3}, and {\it W4}, respectively) beginning in January 2010, yielding a catalog of over 747 million individual sources (\href{http://wise2.ipac.caltech.edu/docs/release/allwise/expsup/}{AllWISE}). Two citizen science projects based on {\it WISE} data have been underway for years, hosted by the Zooniverse citizen science community: \href{https://www.zooniverse.org/projects/povich/milky-way-project}{The Milkyway Project} and \href{http://diskdetective.org}{Disk Detective}.

A primary science goal of {\it WISE} has been to discover and characterize cold brown dwarfs in the solar neighborhood. Indeed, {\it WISE} has met this goal, discovering and defining an entirely new spectral class (``Y dwarfs''; \citealt{2011ApJ...743...50C}, \citealt{2012ApJ...753..156K}), discovering the third and fourth closest systems to the sun (\citealt{2013ApJ...767L...1L}, \citealt{2014ApJ...786L..18L}), and even putting constraints on the existence of a distant companion to the sun \citep{2014ApJ...781....4L}.  Two primary methods to find brown dwarfs with {\it WISE} data have emerged: 1) identifying objects with the unique {\it WISE} colors of cold brown dwarfs (e.g., \citealt{2011ApJS..197...19K}) and 2) identifying objects with significant proper motions (e.g., \citealt{2014ApJ...781....4L}, \citealt{2014ApJ...783..122K}, \citealt{2016ApJ...817..112S}, \citealt{2016ApJS..224...36K}). 
Since the high proper motions and parallactic motions of nearby objects create elongated images with degraded signal-to-noise ratios in some catalogs, finding the nearest brown dwarfs and outer solar system objects requires an approach tuned to moving sources (the second method).

However, so far, searches for high proper motion brown dwarfs in {\it WISE} data have yet to take advantage of the full depth of the {\it WISE} images.  The proper motion surveys of \cite{2014ApJ...781....4L} and \cite{2016ApJ...817..112S} used the {\it WISE} single exposure images (i.e., not coadded), which are limited to a {\it W1} magnitude of $\sim$\,15.3 and a {\it W2} magnitude of $\sim$\,14.5.  The proper motion surveys of \cite{2014ApJ...783..122K} and \cite{2016ApJS..224...36K} were generally limited to objects with {\it W1} magnitudes less than 14 because of criteria imposed to select objects with real proper motions.  In contrast, the survey depths of the {\it WISE} All-Sky catalog are approximately 16.6\,mag in {\it W1} and 16.0\,mag in {\it W2} according to the \href{http://wise2.ipac.caltech.edu/docs/release/allsky/expsup/sec2\_2.html}{{\it WISE} All-Sky data}, indicating that there is a wealth of unexplored proper motion space in the {\it WISE} data. Furthermore, folding in NEOWISE-Reactivation imaging increases the time baseline available in the {\it W1} and {\it W2} bands by a factor $\sim$\,10 relative to that available during AllWISE processing; this time baseline has not been fully exploited.

To search this unexplored space, we have launched a new citizen science project called ``Backyard Worlds: Planet 9''. This project, online at \url{http://backyardworlds.org}, has already engaged roughly 76,000 participants (based on the number of unique IP addresses accessing the site) to examine time-resolved coadds of WISE images. These volunteers have performed more than 3.4 million classifications of these images since the site's launch on February 15, 2017.  Besides seeking new brown dwarfs and nearby stars, the project is a sensitive all-sky search for planets orbiting the Sun beyond Pluto, like the proposed $\sim$\,10\,$M_{\bigoplus}$ object in a $\sim 600$\,au orbit called Planet Nine \citep{2016AJ....151...22B, 2016ApJ...824L..25F}.  

We report here on the first significant discovery by the Backyard Worlds: Planet 9 project: a new brown dwarf in the solar neighborhood, WISEA J110125.95$+$540052.8, hereafter WISEA 1101+5400, identified as a candidate from volunteers just six days  after the project's launch. We describe the citizen science methodology, show a low resolution spectrum of the new T5.5 dwarf, and discuss this discovery as an indicator of the project's search sensitivity.

%-------------------------------------------------%
\section{The Backyard Worlds: Planet 9 Citizen Science Project}
\label{sec:backyardworlds}

After the end of the {\it WISE} cryogenic mission in 2010, the telescope continued to survey the sky in the {\it W1} and {\it W2} bands via the NEOWISE enhancement to the WISE data processing system \citep{2011ApJ...731...53M} until early February 2011. Then, after a period of hibernation, {\it WISE} began to survey the sky again in December 2013, and continued to survey the sky through the end of 2016, a survey know as NEOWISE-Reactivation (NEOWISE; \citealt{2014ApJ...792...30M}). 

\citet{2017AJ....153...38M} have combined all {\it W1} and {\it W2} exposures from the pre-hibernation and first-year NEOWISE-Reactivation data releases into a set of time-resolved coadds, employing an adaptation of the unWISE image coaddition framework \citep{2014AJ....147..108L}. These coadds typically span a $\sim$4.5 year time baseline, from early 2010 to mid-2014. These coadds enable detection of faint moving objects to a depth of $\gtrsim$1 magnitude below the single-exposure limit, provided the motion is slower than  $\sim$\,6\,$''$ per day. The coadds resemble the {\it WISE} Preliminary Release Atlas stacks in terms of depth of coverage and sensitivity, corresponding to limiting magnitudes of $W1$ = 16.5, $W2$ = 15.5 Vega mag \citep{2011wise.rept....1C}.

On the \href{http://www.backyardworlds.org}{Backyard Worlds: Planet 9 website,} we show users four of these \citet{2017AJ....153...38M} coadds, after some additional manipulation to highlight moving objects (described below). The unWISE coaddition divides the sky into 18,240 $1^{\circ}.56 \times 1^{\circ}.56$ astrometric footprints, called "tiles".  We divided each of these tiles into 64 subtiles, 256 $\times$ 256 pixels in size, to facilitate examination on a laptop or cell phone screen.  To highlight moving objects in this data set, we created four difference images to display online:
\begin{eqnarray*}
\rm difference \ image \ 1 = epoch 1 - median(epoch 3, epoch 4) \\
\rm difference \ image \ 2 = epoch 2 - median(epoch 3, epoch 4) \\
\rm difference \ image \ 3 = epoch 3 - median(epoch 1, epoch 2) \\
\rm difference \ image \ 4 = epoch 4 - median(epoch 1, epoch 2) \\
\end{eqnarray*}
We limited each difference image to the range $-0.5 \sigma$ to $9.5 \sigma$, where $\sigma$ represents a typical standard deviation for the noise in each band ($\sigma = 2.1$ Vega nanomaggies for {\it W1}, and $\sigma = 6.9$ Vega nanomaggies for {\it W2}).  We then created a smoothed version of each of the limited difference images above using a 12$\times$12 tophat kernel, and subtracted it from each difference to remove smoothly-varying galactic backgrounds and some electronic noise. The resulting difference images come close to reaching the Poisson noise floor (within a factor of $\sim$$\sqrt{2}$ from the differencing), but with residual subtraction artifacts and electronic noise stripes dominating some regions of most images.

We then applied a standard $\arcsinh{}$ stretch, replacing each difference image with $\arcsinh{(10.0*\rm{image})}$. We constructed our final false-color images by assigning the {\it W1} difference image to the blue channel, the {\it W2} difference image to the red channel, and the mean of the {\it W1} and {\it W2} images to the green channel. This processing yields a color scale where the maximum values correspond roughly to the peaks of point sources with {\it W1} = 16.3 mag and {\it W2} = 15.0 mag. 

We constructed the citizen science website ``Backyard Worlds: Planet 9'' using the Zooniverse project builder platform, at \url{http://www.zooniverse.org/lab}. We uploaded the sets of four difference images to the site to be viewed by users as animated "flipbooks".  The classification task consists of viewing one flipbook and searching it for candidate moving objects, then marking any such objects in all four images using a marking tool. A tutorial and a "Field Guide" provide examples of two different kinds of candidate moving objects, called ``movers'' and ``dipoles''.  A "mover" is a fast-moving source which travels an angular distance greater than the {\it W1}/{\it W2} FWHM over the course of our $\sim$\,4.5 year time baseline. A "dipole" is simply a slower-moving source that moves less than one FWHM in total, so it appears as a negative image right next to a positive image. To facilitate the search, the site features buttons that allow users to zoom in on the flipbook images and also to invert the colors in the flipbook images as an aid to colorblind users.

The primary challenge for users is distinguishing movers and dipoles from various artifacts, the most common of which are stars and galaxies that have been partially-subtracted because of variability or small alignment errors. Cosmic ray hits, optical ghosts and latent images are also common \citep{2012wise.rept....1C}. The online tutorial and the "Field Guide" provide examples of these common artifacts. 

The site provides two other options for users to report their findings to the science team besides marking flipbooks with the marking tool.  All Zooniverse projects are tied to a bulletin board-style social network called ``TALK'' where users can flag interesting objects with \#dipole and \#mover hashtags. Additionally, the site instructs advanced users to research potentially interesting moving sources with both the \href{simbad.u-strasbg.fr/simbad/sim-fcoo}{SIMBAD} and \href{vizier.u-strasbg.fr}{VizieR} databases, and to report objects that do not appear in the SIMBAD database using \href{https://docs.google.com/forms/d/e/1FAIpQLSfg-JRaoTQ9V0miAmH05kBDole9O_pUmIVKxXXBd9FlA77D-g/viewform?c=0&w=1}{a web form}.

\begin{figure*}
\centering
\includegraphics[width=7.5in, clip, trim=1.5cm 4cm 0cm 4cm]{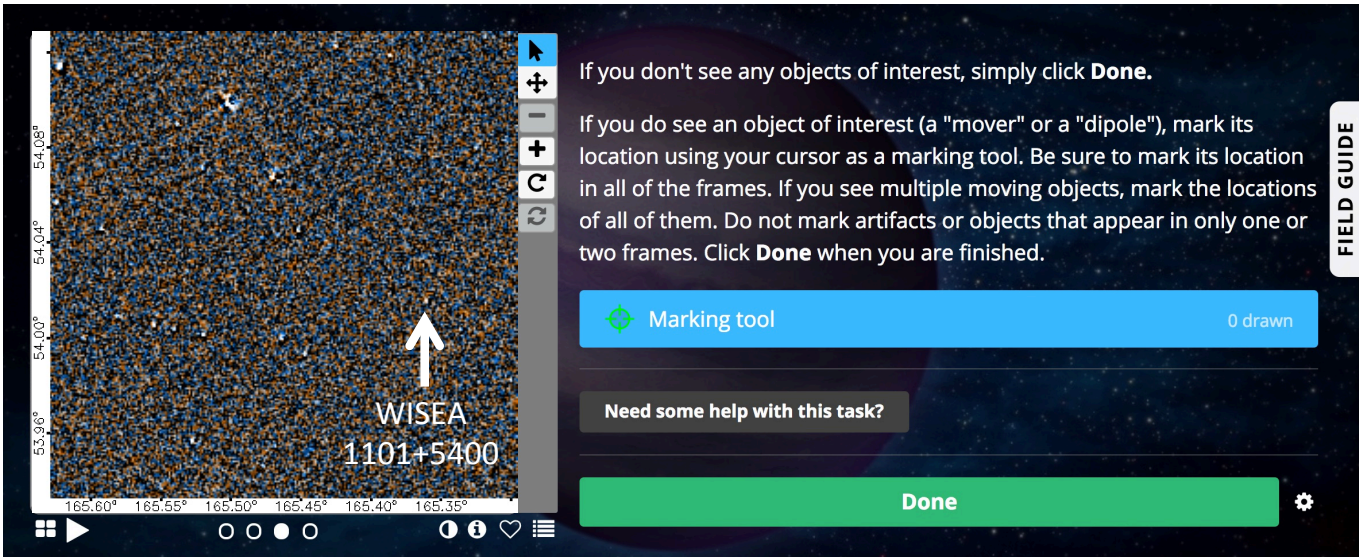}
\centering
\caption{The \url{backyardworlds.org} classification interface showing one frame of the flipbook in which  WISEA 1101+5400 was discovered.  The brown dwarf appears as a faint "dipole" at the tip of the white arrow.  The other sources in the image are subtraction artifacts. Poisson noise and diagonal striping from electronic noise are also apparent. Distinguishing the brown dwarf from the artifacts requires viewing all four difference images in the flipbook by clicking the "play" button in the lower left or by clicking on the four radio buttons under the center of the flipbook.} 
\label{fig:flipbook}
\end{figure*}

Figure \ref{fig:flipbook} shows the \url{backyardworlds.org} classification interface. On the left is the flipbook, showing three dominant noise components; Poisson noise, electronic noise (the crosshatching), and subtraction artifacts (like the source at $\alpha = 165.51^{\circ}, \delta=54.10^{\circ}$).  On the right is the classification question: ``If you don't see any objects of interest...''.   When a user hovers over the flipbook, his or her cursor becomes a marking tool, and clicking on the flipbook leaves a mark in the form of a green crosshair and records the coordinates of the mark in a file of classification data when the user clicks the blue ``Done'' button.

%-------------------------------------------------%
\section{Discovery of WISEA 1101+5400}
\label{sec:WISEA1101+5400}
%-------------------------------------------------%
On February 21, six days after the launch of Backyard Worlds: Planet 9, a user on the Zooniverse TALK social network noted the existence of a ``small dipole'' in the flipbook showing the subtile centered at R.A. 165.46 degrees, declination 54.03 degrees. On February 23, another user reported this dipole to the science team using the \href{https://docs.google.com/forms/d/e/1FAIpQLSfg-JRaoTQ9V0miAmH05kBDole9O_pUmIVKxXXBd9FlA77D-g/viewform?c=0&w=1}{web form,} noting that a VizieR search turned up a source in the AllWISE catalog with no published spectral type.  Three other users also helped classify this flipbook and noted this object.  This source was the 117th submitted to the web form.  It stood out from the others right away because it looked initially like a WISE-only source, without counterparts in the Digitized Sky Survey images and the 2MASS point source catalog. {\bf (However the object was detected in the 2MASS reject catalog in $J$ band only; see Table~1).} The flipbook in which the source was discovered can be viewed at \url{https://www.zooniverse.org/projects/marckuchner/backyard-worlds-planet-9/talk/subjects/5566284}. 

\begin{figure*}
\centering
\includegraphics[width=7.5in]{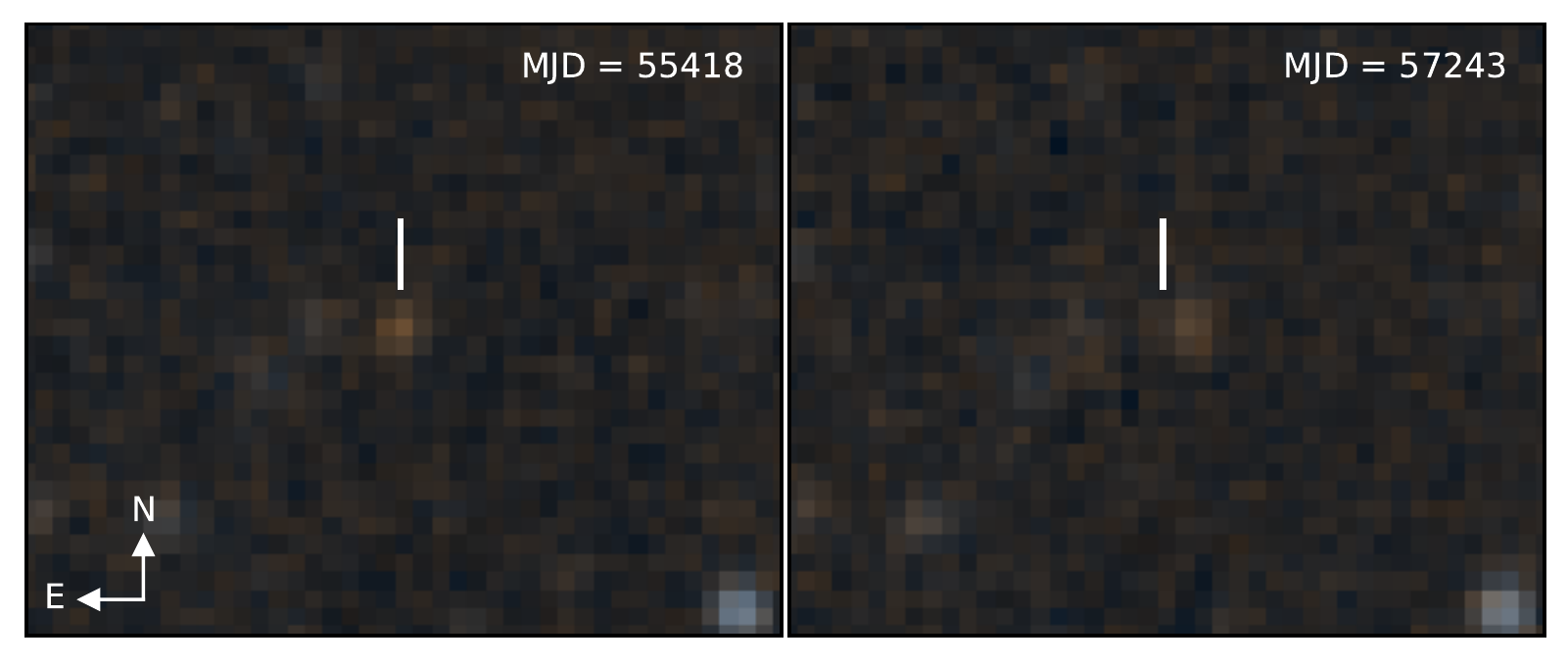}
\centering
\caption{Full-resolution unWISE images of WISEA 1101+5400: a coadd of all pre-hibernation {\it WISE} data (left), and a coadd of all second-year NEOWISE-Reactivation data (right). The white vertical line is fixed to highlight the motion of the brown dwarf during the $\sim$5 year interval. The region shown is $2.0' \times 1.6'$ in size.}
\label{fig:motion}
\end{figure*}

WISEA 1101+5400 appears as a faint "dipole" at the tip of the white arrow in Figure~\ref{fig:flipbook}, where it is barely visible. Figure \ref{fig:motion} shows a clearer view: unWISE images of WISEA 1101+5400 constructed from a coadd of all pre-hibernation {\it WISE} data (left), and a coadd of all second-year NEOWISE-Reactivation data (right). In the time-resolved coadds spanning both pre- and post reactivation {\it WISE} imaging, the motion is visually obvious, whereas this source would be undetected in single exposures and would not display highly significant motion within the pre-hibernation time frame alone. Table \ref{tab:properties} summarizes the basic parameters of WISEA 1101+5400, discussed below.  Besides AllWISE, the object was also detected in Pan-STARRS y-band as PSO J110125.687+540053.435 with a $y$-band PSF magnitude of $\sim$\,19.7.

\begin{deluxetable}{lcc}
\tablehead{Parameter & Value & Ref}
\tablecaption{Properties of WISEA+5400 \label{table:properties}} 
\startdata
R.A. (J2000) & 11:01:25.95 & 1 \\
Decl. (J2000) & +54:00:52.8 & 1\\
$\mu_\alpha \cos \delta$ (\asyr) & $-0.719 \pm 0.010$ & 2\\
$\mu_\delta$ (\asyr) & $0.056 \pm 0.023$ & 2 \\
Pan-STARRS $y_\text{PS}$ & $19.715\pm 0.113$ & 3\\
2MASS $J$ & $17.309 \pm 0.205$ & 4\\
2MASS $J$ & $17.308\pm 0.350$ & 2\\
2MASS $H$ & $17.104\pm 0.191$ & 2\\
2MASS $K_{\rm S}$ & $17.112\pm 0.257$ & 2\\
WISE $W1$ & $17.124\pm 0.118$ &	1\\
WISE $W2$ & $15.371 \pm 0.087$ & 1\\
WISE $W3$ & $> 12.718$	& 1\\
WISE $W4$ & $> 8.934$ & 1\\
$y_\text{PS}-J$ & $2.406\pm 0.234$ & 2 \\
$J-W1$ & $0.185\pm 0.237$ & 2\\
$J-W2$ & $1.938\pm 0.223$ & 2\\
$W1-W2$ & $1.753\pm 0.147$ & 2\\
Spectral type & T5.5$\pm 0.5$ & 2\\
Spectroscopic distance (pc) & $34 \pm 5$ & 2\\
\enddata
\tablenotetext{1}{AllWISE}
\tablenotetext{2}{This work. Synthetic magnitudes from the NIR spectrum.}
\tablenotetext{3}{Pan-STARRS}
\tablenotetext{4}{Taken from the 2MASS Point Source `Reject' Table; No detection in the 2MASS PSC.}
\label{tab:properties}
\end{deluxetable}

%-------------------------------------------------%
\section{Characterization of WISEA 1101+5400}
\label{sec:Characterization}
%-------------------------------------------------%
We obtained a low-resolution near-infrared spectrum for WISEA 1101+5400 on 2017 March 19 using the SpeX spectrograph \citep{2003PASP..115..362R} mounted on the 3\,m NASA Infrared Telescope Facility (IRTF).  We used the spectrograph in prism mode with the 0$\farcs$8 slit aligned to the parallactic angle. This resulted in $R~\equiv~\lambda$ / $\Delta\lambda~\approx$~100 spectral data over the wavelength range 0.7--2.5\,$\mu$m. Conditions were clear although the humidity was high at nearly 80\%. The seeing was 0$\farcs$9 at $K$. Given the faint estimated near-infrared magnitude of the source we began with 120\,s AB subtracted exposures in the guide camera using the $H$+$K$ nod filter. Once we had confidently identified the source, we placed it on the slit and then used off-axis guiding with TCS corrections turned off on the IRTF guide camera. 

We obtained 16 individual exposures of 200\,s each in an ABBA dither pattern along the slit. Immediately after the science observations, we observed the A0V-type star HD~99966 at a similar airmass for telluric corrections and flux calibration. Internal flat-field and Ar arc lamp exposures were acquired for pixel response and wavelength calibration, respectively. All data were reduced using the SpeXtool package version 4.1 \citep{2004PASP..116..362C,2003PASP..115..389V} using standard settings.

%-------------------------------------------------%

\begin{figure*}
\centering
\includegraphics[width=7.5in]{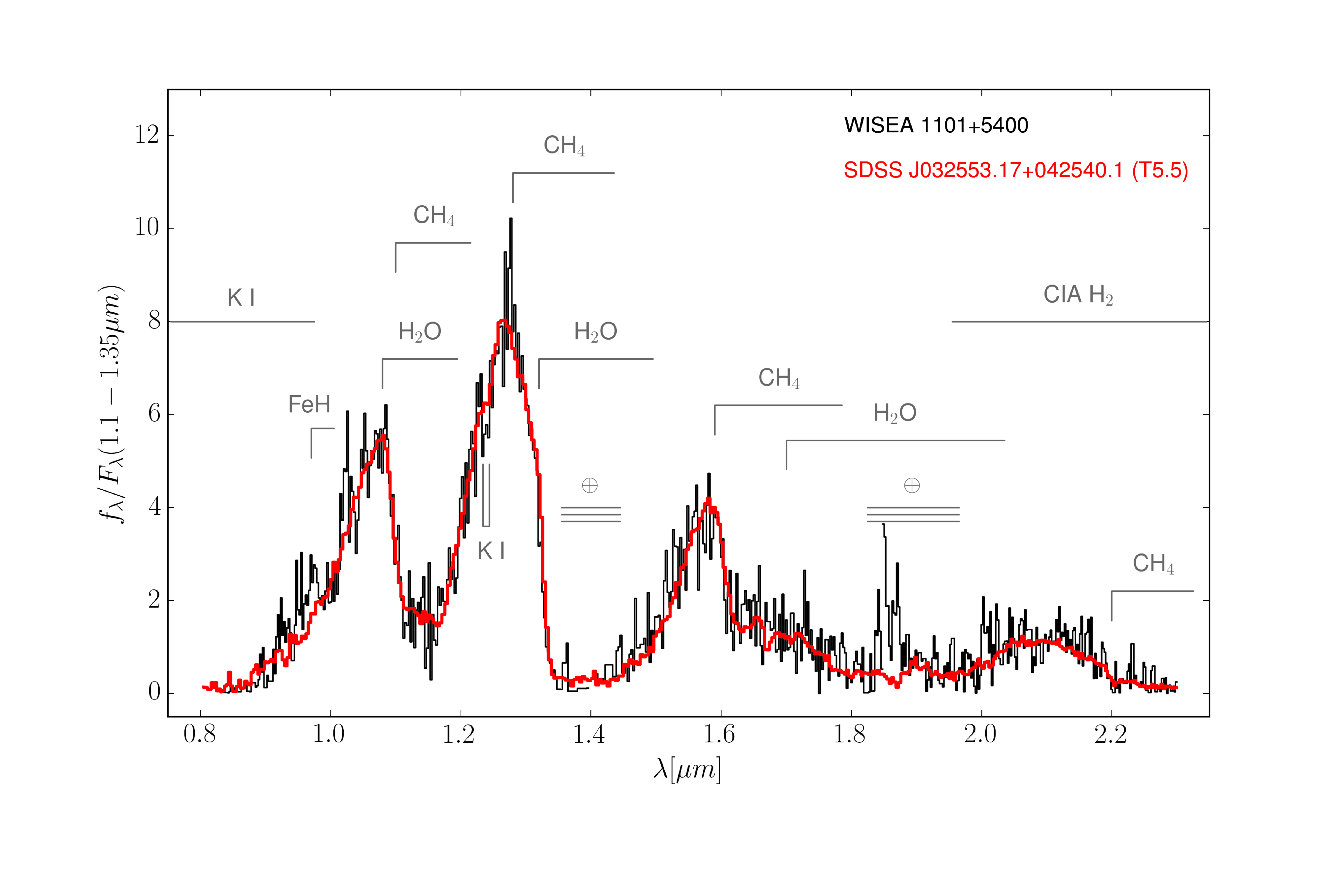}
\centering
\caption{Our SPEX/IRTF spectrum of WISEA 1101+5400 (black) compared to that of SDSS J0325+0425, a field T5.5 dwarf (red). The spectra are normalized in $J$ band. Atomic and molecular opacity sources that define the T dwarf spectral class are indicated.}
\label{fig:spectrum}
\end{figure*}

To determine a spectral type, we performed a goodness-of-fit ($G_k$) statistical test \citep{2008ApJ...678.1372C} of 113 T0--T9 dwarf IRTF Prism spectra from \citet{2013ApJS..205....6M} and \citet{2015ApJ...810..158F} to the near-infrared spectrum from IRTF. The local minimum of the 3rd-order polynomial that describes $G_k(\text{SpT})$ is at T5.5. The best individual match is SDSS J032553.17$+$042540.1, a field T5.5 dwarf. The results of this fitting routine agree well with spectral type estimates determined from the source's {\it W1-W2} color compared to that of the spectral type versus {\it W1-W2} relation derived from the same T dwarf sample. We thus adopt an infrared spectral type of T$5.5 \pm 0.5$. Figure~\ref{fig:spectrum} shows our IRTF/SpeX spectrum of WISEA 1101$+$5400 (black) compared to that of SDSS J0325$+$0425 (red). The spectra are normalized in $J$ band. 

The spectrum for WISEA 1101$+$5400 is moderate to low signal.  Even still, we find no visual peculiarities indicating a deviant gravity, metallicity or atmosphere. By normalizing the observed spectrum to the $y_\text{PS}$ and $J$ magnitudes, we determined synthetic magnitudes in 2MASS $J$, $H$ and $Ks$ bands (provided in Table 1). The spectrum, synthetic near-infrared colors, and measured mid infrared colors are all consistent with a field mid T dwarf. In the absence of a parallax measurement, we used the spectrophotometric relations in $J$, $W1$, and $W2$ for field brown dwarfs from \citet{2016ApJS..225...10F} to estimate a distance of $34\pm 5$\,pc.  A Monte Carlo fitting of the proper motion using all three catalog epochs (2MASS-Reject, Pan-STARRS and AllWISE) yields $\mu_\alpha \cos \delta = -718.7 \pm 9.9$\,\masyr, $\mu_\delta = 56.2 \pm 23.2$\,\masyr. We ran the spatial and kinematic data of the target through the Bayesian Analysis for Nearby Young AssociatioNs (BANYAN) II tool \citep{2013ApJ...762...88M,2014ApJ...783..121G} to assess membership probabilities in nearby young moving groups and found no significant matches.  

%-------------------------------------------------%

\section{Discussion}
\label{sec:Discussion}

\begin{figure*}
\centering
\includegraphics[width=5.5in]{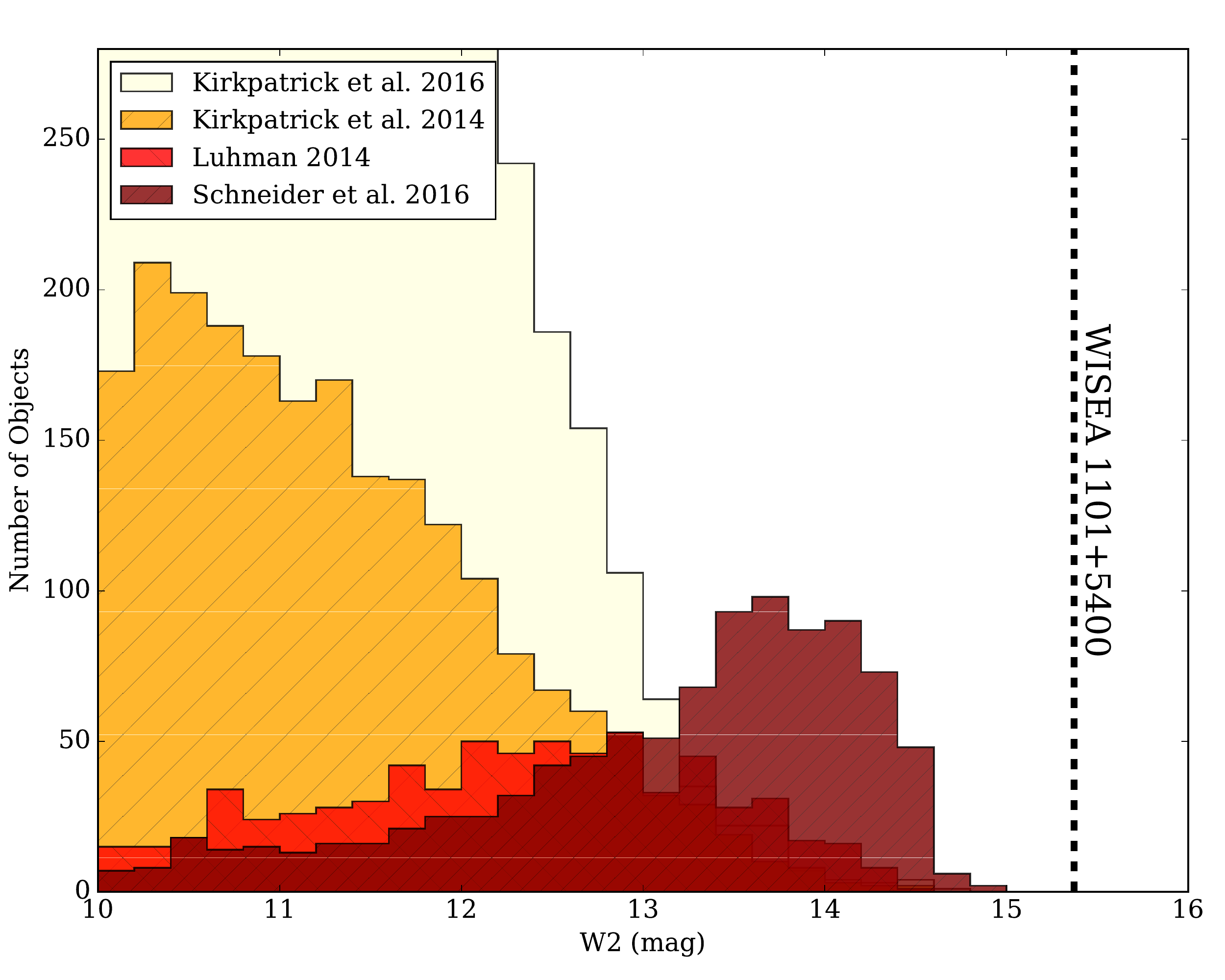}
\centering
\caption{Histograms of the objects discovered by four recent surveys for high proper motion objects with {\it WISE} compared to the $W2$ magnitude of WISEA 1101+5400.}
\label{fig:comparison}
\end{figure*}

The discovery of this object ($W1=17.12$, $W2=15.37$) at $\sim 0.9$ magnitudes below the $W2$ single-exposure detection limit bodes well for the abilities of the citizen scientists at Backyard Worlds: Planet 9 to find additional brown dwarfs and/or distant companions to the Sun. Figure~\ref{fig:comparison} shows the $W2$ magnitude of WISEA 1101$+$5400 compared to histograms of the objects discovered by recent surveys for high proper motion objects using {\it WISE} data by \cite{2014ApJ...781....4L, 2014ApJ...783..122K, 2016ApJS..224...36K} and \cite{2016ApJ...817..112S}. As mentioned above, the proper motion surveys of  \cite{2014ApJ...781....4L} and \cite{2016ApJ...817..112S} used {\it WISE} single exposures and thus have a limiting magnitude of $\sim$14.5 in $W2$.  While not restricted by the single exposure magnitude limit, selection criteria used to pick out high-quality proper motion candidates limited the discoveries from the AllWISE motion surveys of  \cite{2014ApJ...783..122K} and \cite{2016ApJS..224...36K} to $W2$ magnitudes $<$14.5.  WISEA 1101$+$5400 is almost a full magnitude fainter than the faintest discoveries found with any of these previous {\it WISE} proper motion surveys.

By making the conservative assumption that WISEA~1101+5400 is located at the $W2$ limit of this survey and at the minimum detectable proper motion, we determined limiting distances for Backyard Worlds: Planet 9 for each integer spectral type using the spectral type--absolute $W2$ relation of \cite{2012ApJS..201...19D}. We scaled the updated 8\,pc sample of \cite{2012ApJ...753..156K} to these limiting distances and used the Besan\c{c}on Galactic model \citep{2012A&A...538A.106R} to estimate the fraction of objects with proper motions larger than that of WISEA~1101+5400 within each volume. Combining these with Poisson statistics allowed us to estimate the yield of the Backyard Worlds: Planet 9 project: a total of $103_{-29}^{+36}$ L dwarfs, $194_{-33}^{+38}$ T dwarfs, and $18_{-5}^{+7}$ Y dwarfs, including those already known.

\textbf{We estimated the number of known brown dwarfs that we will rediscover by counting up the fraction of known brown dwarfs in each spectral type bin that have measured proper motions larger than that of WISEA~1101+5400 and $W2$ magnitudes brighter than WISEA~1101+5400\footnote{Based on the \cite{2015ApJS..219...33G} compilation available at \url{www.astro.umontreal.ca/~gagne/listLTYs.php}}. 
Then, for each spectral type bin, we multiplied this fraction by the total number of known brown dwarfs, including those without W2 or proper motion measurements. This product represents the number of known brown dwarfs we will rediscover under the conservative assumption that those with missing measurements are distributed in $W2$ and proper motion just like those with measured $W2$ and proper motion. This calculation yields predicted rediscoveries of 59 L dwarfs, 111 T dwarfs and 11 Y dwarfs. Taking these rediscoveries into account, we find that Backyard Worlds: Planet 9 has the potential to discover $37_{-21}^{+39}$ new L dwarfs, $77_{-31}^{+41}$ new T dwarfs, and $6_{-4}^{+7}$ new Y dwarfs.}

WISEA~1101+5400 was undetectable in $W1$. But since data from both $W1$ and $W2$ are processed the same way, we might reasonably assume that the project's $W1$ sensitivity will be 0.9 magnitudes below the single exposure magnitude limit in the $W1$ band, i.e., $W1 = 16.2$. For comparison, the \citet{2014ApJ...781....4L} search for distant companions to the Sun had a limiting magnitude of $W2=14.5$. The \citet{2017AJ....153...65M} search for Planet Nine had a limiting magnitude of $W1=16.66$ (90\% completeness) over a $\sim 2000$ square degree region. Depending on the planet's temperature and atmospheric concentration of CH$_{4}$, models by \cite{2016ApJ...824L..25F} suggest that Planet Nine may be as bright as $W1=16.1$ and $W2=17.2$.

%-------------------------------------------------%
\acknowledgments

The Backyard Worlds: Planet 9 team would like to thank the many Zooniverse volunteers who have participated in this project, from providing feedback during the beta review stage to classifying flipbooks to contributing to the discussions on TALK. We would also like to thank the Zooniverse web development team for their work creating and maintaining the Zooniverse platform and the Project Builder tools. 
This research was supported by a grant from the NASA Science Innovation Fund and NASA ADAP grant NNH17AE75I.
This publication makes use of data products from the Wide-field Infrared Survey Explorer, which is a joint project of the University of California, Los Angeles, and the Jet Propulsion Laboratory/California Institute of Technology, funded by the National Aeronautics and Space Administration.
This publication also makes use of data products from the Two Micron All Sky Survey, a joint project of the University of Massachusetts and the Infrared Processing and Analysis Center/California Institute of Technology, and the Pan-STARRS1 Surveys.


\begin{thebibliography}{}
\expandafter\ifx\csname natexlab\endcsname\relax\def\natexlab#1{#1}\fi

\bibitem[Batygin \& Brown(2016)]{2016AJ....151...22B} Batygin, K., \& Brown, M.~E.\ 2016, \aj, 151, 22 

\bibitem[Boyajian et al.(2016)]{2016MNRAS.457.3988B} Boyajian, T.~S., LaCourse, D.~M., Rappaport, S.~A., et al.\ 2016, \mnras, 457, 3988 

\bibitem[{Cushing {et~al.}(2004)Cushing, Vacca, \&
  Rayner}]{2004PASP..116..362C}
Cushing, M.~C., Vacca, W.~D., \& Rayner, J.~T. 2004, The Publications of the
  Astronomical Society of the Pacific, 116, 362

\bibitem[Cushing et al.(2008)]{2008ApJ...678.1372C} Cushing, M.~C., et al. 2008, \apj, 678, 1372

\bibitem[Cushing et al.(2011)]{2011ApJ...743...50C} Cushing, M.~C., Kirkpatrick, J.~D., Gelino, C.~R., et al.\ 2011, \apj, 743, 50 

\bibitem[Cutri et al.(2011)]{2011wise.rept....1C} Cutri, R.~M., Wright, E.~L., Conrow, T., et al.\ 2011, Explanatory Supplement to the WISE Preliminary Data Release Products,  

\bibitem[Cutri et al.(2012)]{2012wise.rept....1C} Cutri, R.~M., Wright, E.~L., Conrow, T., et al.\ 2012, Explanatory Supplement to the WISE All-Sky Data Release Products,  

\bibitem[Dupuy \& Liu(2012)]{2012ApJS..201...19D} Dupuy, T.~J., \& Liu, M.~C.\ 2012, \apjs, 201, 19 

\bibitem[{Faherty} {et~al.}(2016)]{2016ApJS..225...10F} Faherty, J.~K., et al.\ 2016, \apjs, 225, 10

\bibitem[Filippazzo et al.(2015)]{2015ApJ...810..158F} Filippazzo, J.~C., et al.\ 2015, \apj, 810, 158

\bibitem[Fortney et al.(2016)]{2016ApJ...824L..25F} Fortney, J.~J., Marley, M.~S., Laughlin, G., et al.\ 2016, \apjl, 824, L25 

\bibitem[Gagn{\'e} et al.(2014)]{2014ApJ...783..121G} Gagn{\'e}, J., Lafreni{\`e}re, D., Doyon, R., Malo, L., \& Artigau, {\'E}.\ 2014, \apj, 783, 121 

\bibitem[Gagn{\'e} et al.(2015)]{2015ApJS..219...33G} Gagn{\'e}, J., Faherty, J.~K., Cruz, K.~L., et al.\ 2015, \apjs, 219, 33 

\bibitem[Kirkpatrick et al.(2011)]{2011ApJS..197...19K} Kirkpatrick, J.~D., Cushing, M.~C., Gelino, C.~R., et al.\ 2011, \apjs, 197, 19 

\bibitem[Kirkpatrick et al.(2012)]{2012ApJ...753..156K} Kirkpatrick, J.~D., Gelino, C.~R., Cushing, M.~C., et al.\ 2012, \apj, 753, 156 

\bibitem[Kirkpatrick et al.(2014)]{2014ApJ...783..122K} Kirkpatrick, J.~D., Schneider, A., Fajardo-Acosta, S., et al.\ 2014, \apj, 783, 122 

\bibitem[Kirkpatrick et al.(2016)]{2016ApJS..224...36K} Kirkpatrick, J.~D., Kellogg, K., Schneider, A.~C., et al.\ 2016, \apjs, 224, 36 

\bibitem[Kuchner et al.(2016)]{2016ApJ...830...84K} Kuchner, M.~J., Silverberg, S.~M., Bans, A.~S., et al.\ 2016, \apj, 830, 84 

\bibitem[Lang(2014)]{2014AJ....147..108L} Lang, D.\ 2014, \aj, 147, 108 

\bibitem[Lintott et al.(2008)]{2008MNRAS.389.1179L} Lintott, C.~J., Schawinski, K., Slosar, A., et al.\ 2008, \mnras, 389, 1179 

\bibitem[Lintott et al.(2009)]{2009MNRAS.399..129L} Lintott, C.~J., Schawinski, K., Keel, W., et al.\ 2009, \mnras, 399, 129 

\bibitem[Luhman(2013)]{2013ApJ...767L...1L} Luhman, K.~L.\ 2013, \apjl, 767, L1 

\bibitem[Luhman(2014a)]{2014ApJ...781....4L} Luhman, K.~L.\ 2014a, \apj, 781, 4 

\bibitem[Luhman(2014b)]{2014ApJ...786L..18L} Luhman, K.~L.\ 2014b, \apjl, 786, L18 

\bibitem[Mace et al.(2013)]{2013ApJS..205....6M} Mace, G.~N., Kirkpatrick, J.~D., Cushing, M.~C., et al.\ 2013, \apjs, 205, 6 

\bibitem[Mainzer et al.(2011)]{2011ApJ...731...53M} Mainzer, A., Bauer, J., Grav, T., et al.\ 2011, \apj, 731, 53 

\bibitem[Mainzer et al.(2014)]{2014ApJ...792...30M} Mainzer, A., Bauer, J., Cutri, R.~M., et al.\ 2014, \apj, 792, 30 

\bibitem[Meisner et al.(2017a)]{2017AJ....153...38M} Meisner, A.~M., Lang, D., \& Schlegel, D.~J.\ 2017a, \aj, 153, 38 

\bibitem[Meisner et al.(2017b)]{2017AJ....153...65M} Meisner, A.~M., Bromley, B.~C., Nugent, P.~E., et al.\ 2017b, \aj, 153, 65 

\bibitem[Malo et al.(2013)]{2013ApJ...762...88M} Malo, L., et al.\ 2013, \apj, 762, 88

\bibitem[Murphy et al.(2017)]{2017arXiv170304544M} Murphy, S.~J., Mamajek, E.~E., \& Bell, C.~P.~M.\ 2017, arXiv:1703.04544 
 
\bibitem[{Rayner {et~al.}(2003)Rayner, Toomey, Onaka, Denault, Stahlberger,
  Vacca, Cushing, \& Wang}]{2003PASP..115..362R}
Rayner, J.~T., Toomey, D.~W., Onaka, P.~M., {et~al.} 2003, The Publications of
  the Astronomical Society of the Pacific, 115, 362

\bibitem[{Robin {et~al.}(2012)Robin, Marshall, Schultheis, \&
  Reyl{\'e}}]{2012A&A...538A.106R}
Robin, A.~C., Marshall, D.~J., Schultheis, M., \& Reyl{\'e}, C. 2012, A{\&}A, 538, 106

\bibitem[Schneider et al.(2016)]{2016ApJ...817..112S} Schneider, A.~C., Greco, J., Cushing, M.~C., et al.\ 2016, \apj, 817, 112 

\bibitem[Schwamb et al.(2012)]{2012ApJ...754..129S} Schwamb, M.~E., Lintott, C.~J., Fischer, D.~A., et al.\ 2012, \apj, 754, 129 

\bibitem[Schwamb et al.(2013)]{2013ApJ...768..127S} Schwamb, M.~E., Orosz, J.~A., Carter, J.~A., et al.\ 2013, \apj, 768, 127 

\bibitem[Silverberg et al.(2016)]{2016ApJ...830L..28S} Silverberg, S.~M., Kuchner, M.~J., Wisniewski, J.~P., et al.\ 2016, \apjl, 830, L28 

\bibitem[{Vacca {et~al.}(2003)Vacca, Cushing, \& Rayner}]{2003PASP..115..389V}
Vacca, W.~D., Cushing, M.~C., \& Rayner, J.~T. 2003, The Publications of the
  Astronomical Society of the Pacific, 115, 389

\bibitem[Wright et al.(2010)]{2010AJ....140.1868W} Wright, E.~L., Eisenhardt, P.~R.~M., Mainzer, A.~K., et al.\ 2010, \aj, 140, 1868-1881 

\end{thebibliography}
\end{document}